\documentclass[aps,pra,showpacs,graphics,twocolumn,floatfix,mathbbm,a4paper]{revtex4-1}
\usepackage{amsmath,amsfonts,amssymb,graphics,graphicx,epsfig,color,times,bbm}
\usepackage{graphicx}
\newtheorem{theorem}{Theorem}

\begin{document}
\bibliographystyle{apsrev}

\newcommand{\Tr}{\text{Tr}}
\newcommand{\proofend}{\hfill\fbox\\\medskip }

%---------------------------------------------------------------------------
\title{Minimal state-dependent proof of measurement contextuality for a qubit}

\author{Ravi Kunjwal}
\email{rkunj@imsc.res.in} 
\affiliation{Optics \& Quantum Information Group,
The Institute of Mathematical
  Sciences, C.I.T Campus, Tharamani, Chennai 600 113, India.}

\author{Sibasish Ghosh}
\email{sibasish@imsc.res.in} 
\affiliation{Optics \& Quantum Information Group, The Institute of Mathematical
  Sciences, C.I.T Campus, Tharamani, Chennai 600 113, India.}
\date{\today}

\begin{abstract}
We show that three unsharp binary qubit measurements are enough to violate a generalized noncontextuality inequality, the LSW inequality, 
in a state-dependent manner. For the case of trine spin axes we calculate the optimal quantum violation of this inequality. Besides, we show that unsharp qubit
measurements do not allow a state-independent violation of this inequality. We thus provide a minimal state-dependent proof of measurement contextuality requiring
one qubit and three unsharp measurements. Our result rules out generalized noncontextual models of these measurements which were previously conjectured to exist. 
More importantly, this class of generalized noncontextual models includes the traditional Kochen-Specker (KS) noncontextual models as a proper subset,
so our result rules out a larger class of models than those ruled out by a violation of the corresponding KS-inequality in this scenario. 
\end{abstract}


\maketitle

\section{Introduction}
Quantum theory does not admit Bell-local or Kochen-Specker-noncontextual hidden variable (KS-NCHV) models. This is manifest in Bell-nonlocality \cite{EPR, Bell64} 
and KS-contextuality \cite{Spe60,KS67}. Both these features arise---at a mathematical level---from the lack of a global joint probability distribution over measurement
outcomes that recovers the measurement statistics predicted by quantum theory. Traditionally, KS-contextuality has been shown
for KS-NCHV models of projective measurements for Hilbert spaces of dimension three or greater \cite{KS67, peres, Clifton, Mermin2theorems, Cab1,Cab2, KRK, KCBS, Oh}.

\begin{figure}
 \includegraphics[width=0.28\textwidth]{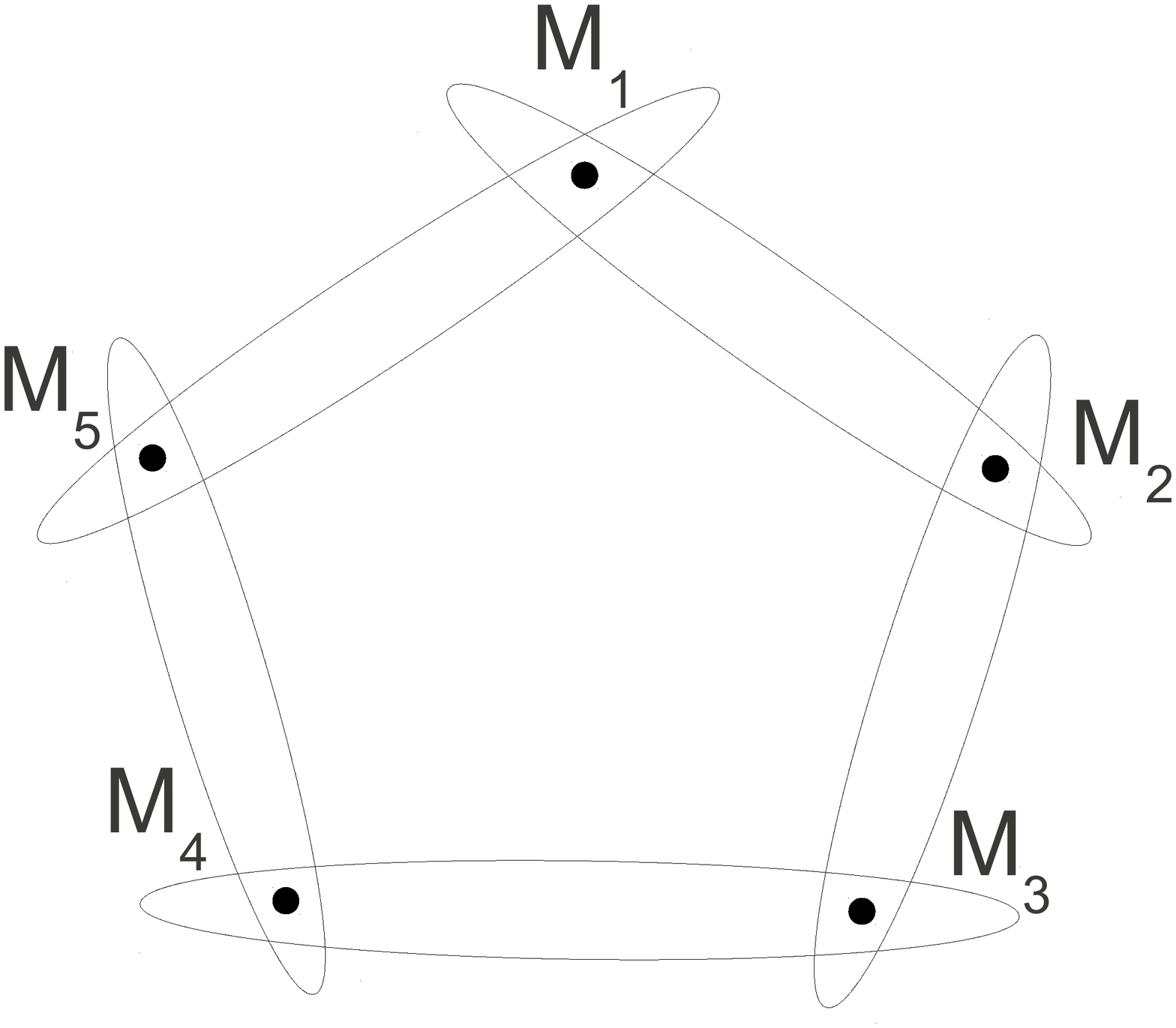}
 \caption{The KCBS \cite{KCBS} contextuality scenario. The vertices represent the measurements and edges represent jointly measurable contexts.}
 \label{kcbs}
 \end{figure}

For projective measurements, KS-noncontextuality assumes that the outcome of a measurement $A$ is independent of whether it is performed together with a measurement
$B$, where $[A,B]=0$, or with measurement $C$, where $[A,C]=0$ and $B$ and $C$ are not compatible, i.e., $[B,C]\neq0$. $B$ and $C$ provide contexts for measurement of $A$. 
A qubit cannot yield a proof of KS-contextuality because it does not admit such a triple of projective measurements. While a
state-independent proof of KS-contextuality holds for any state-preparation, a state-dependent proof requires a special choice 
of the prepared state. The minimal state-independent proof of KS-contextuality requires a qutrit and $13$ projectors \cite{Oh, cabmin}.
The minimal state-dependent proof \cite{KCBS, KRK}, first given by Klyachko et al., requires a qutrit and five projectors (Fig.~\ref{kcbs}). Thus a qutrit is the simplest quantum system that 
allows a proof of KS-contextuality, both state-independent and state-dependent. However, we note that generalizations of KS-noncontextuality for a qubit have been considered earlier \cite{Cab3, busch, caves}
in a manner that is different from our approach. The precise difference, and the merits of our approach over earlier attempts, are discussed at length in Ref. \cite{noODUM}. Here we simply note that
our approach builds upon the work of Spekkens \cite{genNC} and Liang et. al. \cite{OS}, and we consider generalized noncontextuality 
proposed by Spekkens as the appropriate notion of noncontextuality for unsharp measurements \cite{noODUM,genNC}. Generalized noncontextuality allows outcome-indeterministic response functions for unsharp 
measurements in the ontological model while KS-NCHV models insist on outcome-deterministic response functions. In particular,
generalized noncontextuality insists on the noncontextuality of \emph{probability} assignments to measurement outcomes rather than the stronger 
KS-noncontextual assumption of the noncontextuality of \emph{value} assignments. A KS-NCHV model is necessarily generalized-noncontextual but the converse is not true.
Further, while the assumption of KS-noncontextuality applies to projective measurements in quantum theory, the assumption of generalized noncontextuality applies to
all experimental procedures---preparations, transformations, and measurements---in any operational theory.

\begin{figure}
\includegraphics[width=0.28\textwidth]{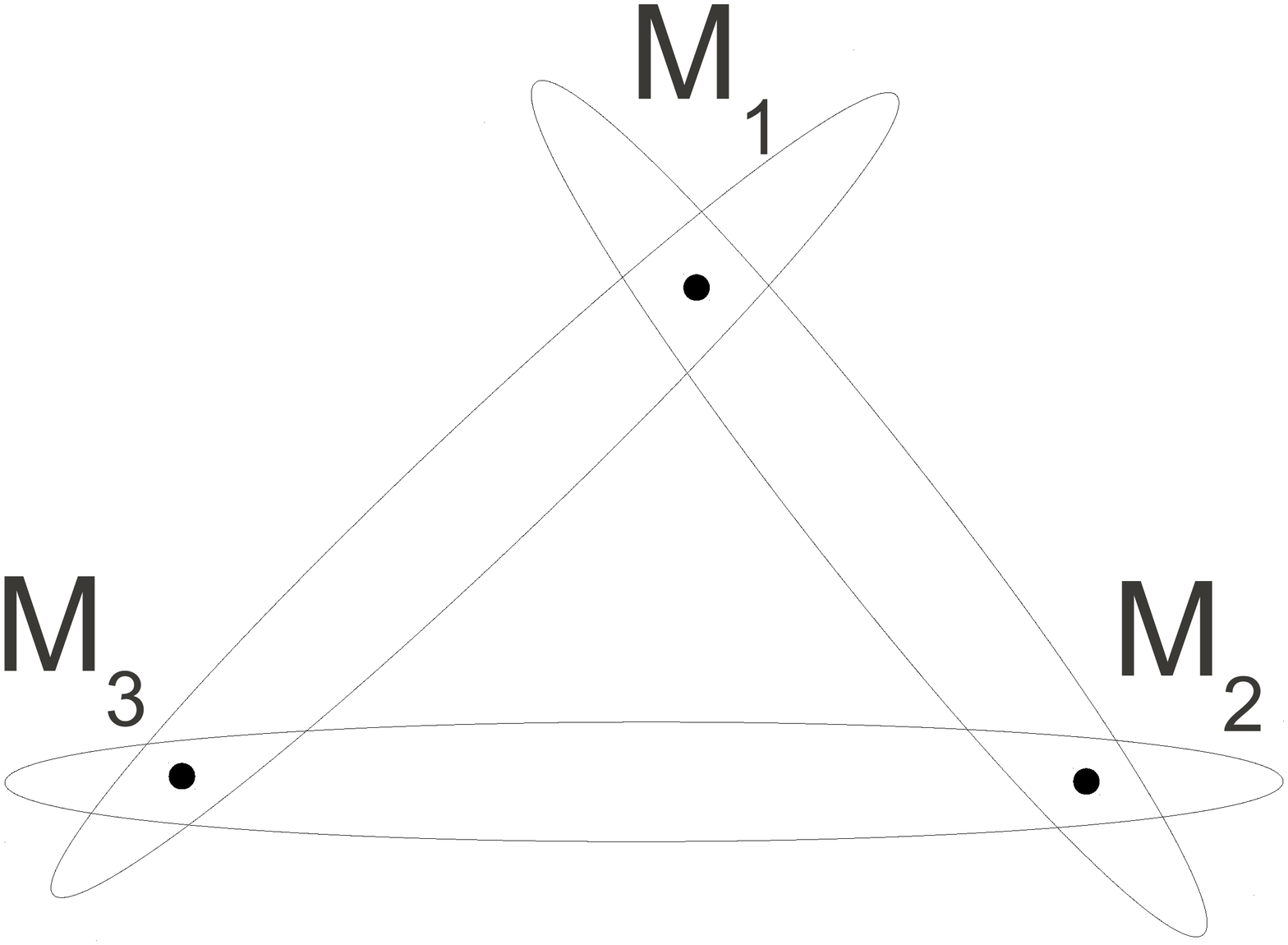}
\caption{Specker's \cite{Spe60} contextuality scenario. The vertices represent the measurements and hyperedges represent jointly measurable contexts.}
\label{specker}
\end{figure}

We define a contextuality scenario as a collection of subsets, called `contexts', of the set of all measurements. A context refers to measurements that can be jointly implemented. 
Conceptually, the simplest possible contextuality scenario, first considered by Specker \cite{Spe60} (Fig.~\ref{specker}), requires three two-valued measurements, $\{M_1,M_2,M_3\}$, to allow for three non-trivial contexts:
$\{\{M_1,M_2\},\{M_1,M_3\},\{M_2,M_3\}\}$. Any other choice of contexts will be trivially KS-noncontextual, e.g., $\{\{M_1,M_2\},\{M_1,M_3\}\}$ is KS-noncontextual because the joint probability distribution
$p(M_1,M_2,M_3)\equiv p(M_1,M_2)p(M_1,M_3)/p(M_1)$ reproduces the marginal statistics. By implication, it is also generalized-noncontextual.
On assigning outcomes $\{+1,-1\}$ noncontextually to the three measurements $\{M_1,M_2,M_3\}$, it becomes obvious that the 
maximum number of anticorrelated contexts possible in a single assignment is two, e.g., for the assignment 
$\{M_1\rightarrow+1,M_2\rightarrow-1,M_3\rightarrow+1\}$, $\{M_1,M_2\}$ and $\{M_2,M_3\}$ are anticorrelated but $\{M_1,M_3\}$ is not.
This puts a KS-noncontextual upper bound of $\frac{2}{3}$ on the probability of anticorrelation when a context is chosen uniformly at random.
Specker's scenario precludes projective measurements because a set of three pairwise commuting projective measurements is trivially jointly measurable and cannot show any contextuality.
One may surmise that it represents a kind of contextuality that is not seen in quantum theory. However, as Liang et al. showed
\cite{OS}, this contextuality scenario can be realized using noisy spin-1/2 observables. If one does not assume
outcome-determinism for unsharp measurements and models them stochastically but noncontextually, then this generalized-noncontextual model
for noisy spin-1/2 observables will obey a bound of $1-\frac{\eta}{3}$, where $\eta \in [0,1]$ is the sharpness associated with each observable.
Formally,
\begin{equation}
 R_3\equiv\frac{1}{3}\sum_{(ij)\in\{(12),(23),(13)\}}\textrm{Pr}(X_i\neq X_j|G_{ij})\leq 1-\frac{\eta}{3},
\end{equation}
where $\textrm{Pr}(X_i\neq X_j|G_{ij})$ is the probability of anticorrelation between the outcomes $X_i, X_j \in \{+1,-1\}$ of measurements $M_i$ and $M_j$,
respectively. $G_{ij}$ denotes the POVM corresponding to the joint implementation of $M_i$ and $M_j$. We will refer to this generalized noncontextuality inequality as the \emph{LSW (Liang-Spekkens-Wiseman) inequality}. This 
is \emph{not} a KS-noncontextual inequality, for which the bound would be $\frac{2}{3}$. A violation of the LSW inequality will rule out generalized noncontextuality and, by implication, 
KS-noncontextuality. A discussion of this generalized noncontextual model and its ontological meaning, compared to the usual KS-NCHV model, is provided in Appendix \ref{modelcompare},
where we also point out the merits of generalized noncontextuality over KS-noncontextuality as a benchmark for nonclassicality. For a more detailed analysis of these issues we refer
the interested reader to Refs. \cite{noODUM,genNC}.

After giving examples of orthogonal and trine spin-axes that did not seem to show a violation of this inequality, Liang et al. left open the question of whether such a violation exists \cite{OS}.
They conjectured that all such triples of POVMs will admit a generalized noncontextual model \cite{genNC}, i.e., the LSW inequality will not be violated.

Our main result is a proof that a state-dependent violation of the LSW inequality is possible. In Section \ref{lswineq} we set up the LSW inequality for 
three unsharp qubit POVMs, in Section \ref{constraintsoneta} we obtain constraints on $\eta$ from joint measurability, and Section \ref{constructjoint} provides
construction of the joint measurement POVMs. In Section \ref{noSI} we prove that noisy spin-1/2 observables do not
allow a state-independent violation of the LSW inequality, followed by our main result in Section \ref{sdviolation}: a state-dependent violation of LSW inequality for the case of trine spin axes.
We conclude with some discussion and open questions in Section \ref{conclude}.
\section{The LSW inequality}\label{lswineq}
The three POVMs considered, $M_k=\{E^k_+, E^k_-\}$, $k\in \{1,2,3\}$, are noisy spin-$\frac{1}{2}$ observables of the form 
\begin{equation}\label{qubitPOVMs}
 E^k_{\pm}\equiv \frac{1}{2}I\pm \frac{\eta}{2}\vec{\sigma}.\hat{n}_k, \quad 0\leq \eta \leq 1.
\end{equation}
That is,
\begin{equation}
 E^k_{\pm}=\frac{1-\eta}{2}I+\eta\Pi^k_{\pm},
\end{equation}
where $\Pi^k_{\pm}=\frac{1}{2}(I\pm\vec{\sigma}.\hat{n}_k)$ are the corresponding projectors. So $E^k_{\pm}$ are noisy versions of the projectors $\Pi^k_{\pm}$, and the observable $\{E^k_+,E^k_-\}$ is therefore a noisy 
(or unsharp) version of the projective measurement $P_k=\{\Pi^k_+,\Pi^k_-\}$ (for $k\in \{1,2,3\}$).
The LSW inequality concerns the following quantity:
\begin{equation}
R_3\equiv\frac{1}{3}\sum_{(ij)\in\{(12),(23),(13)\}}p(X_i\neq X_j|G_{ij}) 
\end{equation}
where $X_i, X_j \in \{+1,-1\}$ label measurement outcomes for $M_i$ and $M_j$, respectively. The joint measurement POVM for the context $\{M_i, M_j\}$ is denoted by $G_{ij}\equiv\{G_{++}^{ij},G_{+-}^{ij},G_{-+}^{ij},G_{--}^{ij}\}$.
$G^{ij}_{X_i,X_j}$ is the joint measurement effect corresponding to the effects $E^i_{X_i}$ and $E^j_{X_j}$, i.e., $\sum_{X_j} G^{ij}_{X_i,X_j}=E^i_{X_i}$, and $\sum_{X_i} G^{ij}_{X_i,X_j}=E^j_{X_j}$.
$R_3$ is the average probability of anticorrelation when one of the three contexts is chosen uniformly at random.

Under a generalized noncontextual model for these noisy spin-1/2 observables, the following bound on $R_3$ holds (cf. \cite{OS}, Section 7.3):
\begin{equation}\label{ncineq}
R_3\leq 1-\frac{\eta}{3} 
\end{equation}
The question is: Does there exist a triple of noisy spin-1/2 observables that will violate this inequality, perhaps for some specific state-preparation? 
\section{Joint measurability constraints on $\eta$}\label{constraintsoneta}
Testing the LSW inequality for a quantum mechanical violation requires a special kind of joint measurability, denoted by jointly measurable contexts $\{\{M_1,M_2\}, \{M_2,M_3\}, \{M_1,M_3\}\}$,i.e.,
pairwise joint measurability but no triplewise joint measurability. This can be achieved by adding noise to projective measurements along three different axes.
For a given choice of $\{\hat{n}_1,\hat{n}_2,\hat{n}_3\}$ in eq. (\ref{qubitPOVMs}), denoting $\hat{n}_i.\hat{n}_j\equiv \cos \theta_{ij}$, a sufficient condition for this kind of joint measurability is
\begin{equation}\label{jointcriterion}
 \eta_l<\eta\leq \eta_u
\end{equation}
where
\begin{equation}\label{lower}
\eta_l=\frac{1}{3} \max_{X_1,X_2,X_3 \in \{\pm1\}} \left\{\sqrt{3+2\sum_{k,l \in \{1,2,3\}, k<l}X_kX_l\cos \theta_{kl}} \right\}
\end{equation}
and 
\begin{equation}\label{upper}
\eta_u=\min_{(ij)\in\{(12),(23),(13)\}}\left\{\frac{1}{\sqrt{1+|\sin \theta_{ij}|}}\right\}
\end{equation}
These are obtained as special cases of the more general joint measurability conditions obtained in Appendix \ref{bounds}, based on Refs. \cite{OS} and \cite{YuOh}. We note that this condition is 
necessary and sufficient for the special case of trine ($\hat{n}_i.\hat{n}_j=-1/2$) and orthogonal ($\hat{n}_i.\hat{n}_j=0$) spin axes.
\section{Joint measurement effects}\label{constructjoint}
We construct the joint measurement POVM, $G_{ij}=\{G^{ij}_{++},G^{ij}_{+-},G^{ij}_{-+},G^{ij}_{--}\}$, such that the given POVMs, $M_i=\{E^i_+,E^i_-\}$ and $M_j=\{E^j_+,E^j_-\}$, are recovered 
as marginals, i.e., $\sum_{X_j} G^{ij}_{X_i,X_j}=E^i_{X_i}$, $\sum_{X_i} G^{ij}_{X_i,X_j}=E^j_{X_j}$, $0\leq G^{ij}_{X_i,X_j}\leq I$, and $\sum_{X_i, X_j} G^{ij}_{X_i,X_j}=I$, where $X_i, X_j \in \{+1,-1\}$.
The joint measurement POVM has the following general form:
\begin{eqnarray}\label{jointbeg}
 G^{ij}_{++}&=&\frac{1}{2}[\frac{\alpha_{ij}}{2}I+\vec{\sigma}.\frac{1}{2}(\eta(\hat{n}_i+\hat{n}_j)-\vec{a}_{ij})]\\
 G^{ij}_{+-}&=&\frac{1}{2}[(1-\frac{\alpha_{ij}}{2})I+\vec{\sigma}.\frac{1}{2}(\eta(\hat{n}_i-\hat{n}_j)+\vec{a}_{ij})]\\
 G^{ij}_{-+}&=&\frac{1}{2}[(1-\frac{\alpha_{ij}}{2})I+\vec{\sigma}.\frac{1}{2}(\eta(-\hat{n}_i+\hat{n}_j)+\vec{a}_{ij})]\\
 G^{ij}_{--}&=&\frac{1}{2}[\frac{\alpha_{ij}}{2}I+\vec{\sigma}.\frac{1}{2}(\eta(-\hat{n}_i-\hat{n}_j)-\vec{a}_{ij})]\label{jointend}
\end{eqnarray}
where $I$ is the $2\times2$ identity matrix, $\vec{\sigma}=(\sigma_x,\sigma_y,\sigma_z)$ are the $2\times2$ Pauli matrices, $\alpha_{ij} \in \mathbb{R}$, and $\vec{a}_{ij} \in \mathbb{R}^3$. The necessary and sufficient conditions for these to be valid 
qubit effects, $0\leq G^{ij}_{X_i,X_j}\leq I$, $\forall X_i,X_j \in \{+1,-1\}$, are equivalent to the following inequalities \cite{heinosaari},
\begin{equation}\label{valid1}
\sqrt{2\eta^2(1+\hat{n}_i.\hat{n}_j)+|\vec{a}_{ij}|^2+2\eta|(\hat{n}_i+\hat{n}_j).\vec{a}_{ij}|}\leq\alpha_{ij}
\end{equation}
\begin{equation}\label{valid2}
\alpha_{ij}\leq2-\sqrt{2\eta^2(1-\hat{n}_i.\hat{n}_j)+|\vec{a}_{ij}|^2+2\eta|(\hat{n}_i-\hat{n}_j).\vec{a}_{ij}|},
\end{equation}
where $\eta_l<\eta\leq \eta_u$. The construction of the joint measurement POVM and derivation of the necessary and sufficient condition for its validity, (\ref{valid1})-(\ref{valid2}),
are provided in Appendix \ref{jointconstruct}. The joint measurement effects corresponding to anticorrelation sum to
\begin{equation}\label{anticorr}
 G^{ij}_{+-}+G^{ij}_{-+}=(1-\frac{\alpha_{ij}}{2})I+\frac{1}{2}\vec{\sigma}.\vec{a}_{ij}.
\end{equation}
\section{No state-independent violation}\label{noSI}
We will now show that no state-independent violation of the LSW inequality with qubit POVMs is possible.
\begin{theorem}
There exists no state-independent violation of the generalized-noncontextual inequality $R_3 \leq 1-\frac{\eta}{3}$ using a triple of qubit POVMs, $M_k\equiv \{E^k_{\pm}\}, k\in\{1,2,3\}$,
that are pairwise jointly measurable but not triplewise jointly measurable. 
\end{theorem}
{\it Proof.---} In quantum theory, the probability $R^Q_3$ for anticorrelation of measurement outcomes for pairwise joint measurements of $M_k\equiv\{E^k_+,E^k_-\}$ (where $k\in\{1,2,3\}$) has the following form
for a qubit state $\rho$:
\begin{equation}
R^Q_3\equiv\frac{1}{3}\sum_{(ij)\in\{(12),(23),(13)\}}\textrm{Tr}(\rho(G^{ij}_{+-}+G^{ij}_{-+})),
\end{equation}
The condition for violation of noncontextual inequality (\ref{ncineq}) is $R_3^Q > 1-\frac{\eta}{3}$. Using (\ref{anticorr}), this reduces to 
\begin{equation}
 \textrm{Tr}\big(\rho\sum_{(ij)}(\alpha_{ij}I-\vec{\sigma}.\vec{a}_{ij})\big)<2\eta
\end{equation}
Using the standard $2 \times 2$ Pauli matrices and $\rho$ parameterized by $0\leq q\leq 1$ and $\hat{n}=(\sin \theta \cos \phi, \sin \theta \sin \phi, \cos \theta)$:
\begin{eqnarray}\label{state1}
 \rho&=&q|\psi\rangle\langle\psi|+(1-q)(I-|\psi\rangle\langle\psi|)\\\label{state2}
 |\psi\rangle&=&\left( \begin{array}{c}
\cos \frac{\theta}{2}\\
e^{i\phi}\sin \frac{\theta}{2}
\end{array}\right)=\cos \frac{\theta}{2}|0\rangle+e^{i\phi}\sin \frac{\theta}{2}|1\rangle
\end{eqnarray}
the condition for violation becomes
\begin{equation}\label{viol}
 \sum_{(ij)}\alpha_{ij}+\lambda_{\rho}<2\eta
\end{equation}
where 
\begin{equation}\label{statedep}
\lambda_{\rho}\equiv(1-2q)\vec{a}.\hat{n} \in [-|\vec{a}|,|\vec{a}|]
\end{equation}
denotes the state-dependent term in the condition and $\vec{a}=(a_x,a_y,a_z)$ is given by
\begin{equation}
 a_x=\sum_{(ij)}(\vec{a}_{ij})_x, \quad a_y=\sum_{(ij)}(\vec{a}_{ij})_y, \quad a_z=\sum_{(ij)}(\vec{a}_{ij})_z.
\end{equation}
For a state-independent violation, either the state-dependent term in (\ref{viol}),
$\lambda_{\rho}$, must vanish for all qubit states $\rho$, or  $\sum_{(ij)}\alpha_{ij}+\max_{\rho}\lambda_{\rho}<2\eta$ should hold. The first case, $\lambda_{\rho}=0 \quad \forall \rho$, 
requires $\vec{a}=0$, since $\vec{a}$ is the only term in $\lambda_{\rho}$ that depends on the joint measurement POVM. 
This means $a_x=a_y=a_z=0$, so that $\lambda_{\rho}=0$ for all $\rho$. The second case requires $\sum_{(ij)}\alpha_{ij}+|\vec{a}|<2\eta$. In both cases,
we have the following lower bound on $\alpha_{ij}$, from inequality (\ref{valid1}):
\begin{equation}
\alpha_{ij}>\sqrt{2}\eta\sqrt{1+\hat{n}_i.\hat{n}_j}  
\end{equation}
Taking the sum of $\alpha_{ij}$, $(ij) \in \{(12),(23),(13)\}$, we have
\begin{equation}
\sum_{(ij)}\alpha_{ij}>\sqrt{2}\eta\sum_{(ij)}\sqrt{1+\hat{n}_i.\hat{n}_j}
\end{equation}
For the first case, the condition for state-independent violation is, $\sum_{(ij)}\alpha_{ij}<2\eta$, while for the second case the condition for such a violation is $\sum_{(ij)}\alpha_{ij}+|\vec{a}|<2\eta$.
Given the lower bound on $\sum_{(ij)}\alpha_{ij}$, it follows that a necessary condition for state-independent violation of the LSW inequality
is:
\begin{equation}
 \sum_{(ij)}\sqrt{1+\hat{n}_i.\hat{n}_j}<\sqrt{2}.
\end{equation}
We will show that there exists no choice of measurement directions that will satisfy this necessary condition, thereby ruling out a state-independent violation of the LSW inequality.
The particular cases of orthogonal axes ($\hat{n}_i.\hat{n}_j=0$) or trine spin axes ($\hat{n}_i.\hat{n}_j=-1/2$), used in \cite{OS}, are clearly ruled out by this necessary condition.
Denoting $\hat{n}_i.\hat{n}_j\equiv\cos\theta_{ij}$, the necessary condition for violation is
\begin{equation}\label{necc}
  |\cos \frac{\theta_{12}}{2}|+|\cos \frac{\theta_{13}}{2}|+|\cos \frac{\theta_{23}}{2}|<1
\end{equation}
Without loss of generality, the three directions can be parameterized as:
\begin{eqnarray}\label{mmts1}
 \hat{n}_1&\equiv&(0,0,1),\\
 \hat{n}_2&\equiv&(\sin \theta_{12},0,\cos \theta_{12}),\\
 \hat{n}_3&\equiv&(\sin \theta_{13} \cos \phi_3, \sin \theta_{13} \sin \phi_3, \cos \theta_{13}).\label{mmts2}
\end{eqnarray}
where
$$0<\frac{\theta_{ij}}{2}<\frac{\pi}{2} \quad \forall (ij) \in \{(12),(13),(23)\},\quad 0\leq \phi_3< 2\pi,$$
and $\cos \theta_{23}=\sin \theta_{12}\sin \theta_{13} \cos \phi_3+\cos \theta_{12}\cos \theta_{13}.$
This implies:
\begin{equation}\label{angles}
\cos(\theta_{12}+\theta_{13})\leq \cos(\theta_{23}) \leq \cos(\theta_{12}-\theta_{13}).
\end{equation}
Then 
\begin{eqnarray}\nonumber
&&\min_{\theta_{12}, \theta_{13},\theta_{23}}\big\{|\cos \frac{\theta_{12}}{2}|+|\cos \frac{\theta_{13}}{2}|+|\cos \frac{\theta_{23}}{2}|\big\}\geq\\\nonumber
&&\min_{\theta_{12},\theta_{13}}\big\{|\cos \frac{\theta_{12}}{2}|+|\cos \frac{\theta_{13}}{2}|+\sqrt{\frac{1+\cos(\theta_{12}+\theta_{13})}{2}}\big\}>1. 
\end{eqnarray}
This contradicts the necessary condition (\ref{necc}). Hence, there is no state-independent violation of the LSW inequality (\ref{ncineq}) allowed by noisy spin-1/2 observables.
\proofend

\section{State-dependent violation of the LSW inequality}\label{sdviolation}
Our main result is that the LSW inequality can be violated in a state-dependent manner.
From the condition for violation (\ref{viol}), it follows that a necessary condition for state-dependent violation is $\sum_{(ij)}\alpha_{ij}-|\vec{a}|<2\eta$.
An optimal choice of $\rho$ that yields $\lambda_{\rho}=-|\vec{a}|$ corresponds to $q=1$ and $\vec{a}.\hat{n}=|\vec{a}|$, i.e.,
$$\cos \theta = \frac{a_z}{|\vec{a}|}, \quad \tan \phi=\frac{a_y}{a_x}.$$

\begin{figure}
\includegraphics[width=0.28\textwidth]{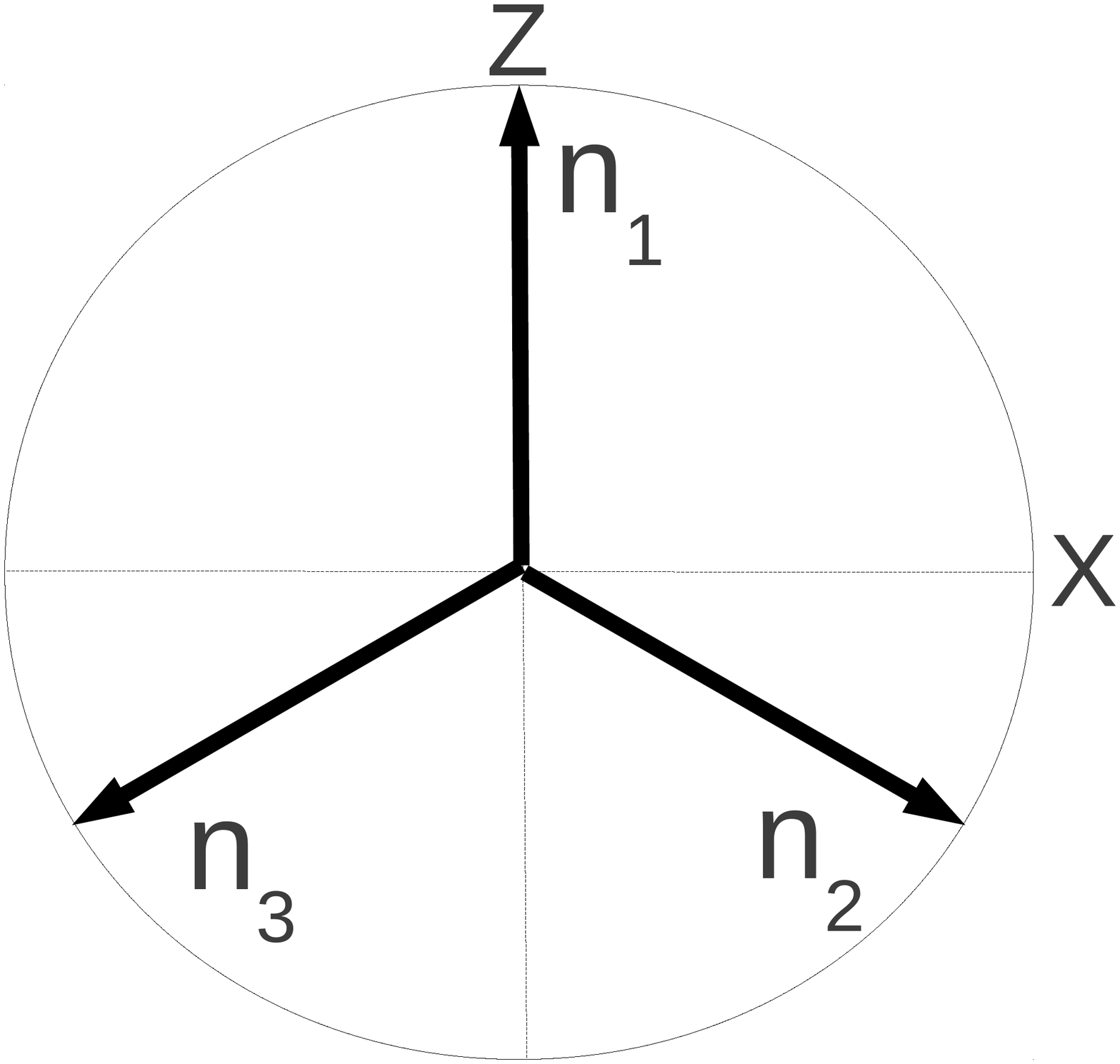}
\caption{Choice of measurement directions $\{\hat{n}_1,\hat{n}_2,\hat{n}_3\}$ along trine spin axes in the Z-X plane.}
\label{plane}
\end{figure}

With this choice of $\rho$ the question becomes: Does there exist a choice of $\{\hat{n}_1,\hat{n}_2,\hat{n}_3\}, \eta, \{\alpha_{ij}, \vec{a}_{ij}\}$ such that
$\sum_{(ij)}\alpha_{ij}-|\vec{a}|<2\eta$? We show that this is indeed the case. We define
\begin{equation}\label{C}
 C\equiv2\eta-(\sum_{(ij)}\alpha_{ij}-|\vec{a}|),
\end{equation}
so that $C>0$ indicates a state-dependent violation. Note that violation of the LSW inequality $R_3^Q\leq 1-\frac{\eta}{3}$ is characterized by
\begin{equation}
 S\equiv R_3^Q-(1-\frac{\eta}{3})=\frac{C}{6}
\end{equation}
where $S>0$ for a state-dependent violation. Given a coplanar choice of $\{\hat{n}_1,\hat{n}_2,\hat{n}_3\}$, and $\eta$ satisfying $\eta_l<\eta\leq\eta_u$, the optimal value of $C$
---denoted as $C^{\{\hat{n}_i\},\eta}_{\max}$---is given by 
\begin{eqnarray}\label{cmax}
 &&C^{\{\hat{n}_i\},\eta}_{\max}=2\eta\nonumber\\&+&\sum_{(ij)}\left(\sqrt{1+\eta^4(\hat{n}_i.\hat{n}_j)^2-2\eta^2}-(1+\eta^2 \hat{n}_i.\hat{n}_j)\right),
\end{eqnarray}
which is obtained in Appendix \ref{optimal}.
We obtain a state-dependent violation of the LSW inequality for trine axes (Fig. \ref{plane}):
\begin{theorem}\label{thm2}
The optimal violation of the LSW inequality for measurements along trine spin axes, i.e., $\{\hat{n}_1,\hat{n}_2,\hat{n}_3\}$ such that $\hat{n}_1.\hat{n}_2=\hat{n}_2.\hat{n}_3=\hat{n}_1.\hat{n}_3=-1/2$, occurs for $|\psi\rangle=\frac{1}{\sqrt{2}}(|0\rangle+i|1\rangle)$ if the plane of
measurements is the ZX plane. The lower and upper bounds on $\eta$ are $\eta_l=\frac{2}{3}\approx0.667$ and $\eta_u=\sqrt{3}-1\approx 0.732$. The joint measurement POVM is given by $\alpha_{ij}=1+\eta^2\hat{n}_i.\hat{n}_j$ and $\vec{a}_{ij}=(0,\sqrt{1+\eta^4(\hat{n}_i.\hat{n}_j)^2-2\eta^2},0)$.
The optimal violation corresponds to $\eta \rightarrow \eta_l$, so that $\alpha_{12}=\alpha_{13}=\alpha_{23}\rightarrow 1-\frac{\eta_l^2}{2}=\frac{7}{9}$, $|\vec{a}_{ij}|\rightarrow \frac{\sqrt{13}}{9} \quad \forall (ij)$,
$C^{trine}_{\max}\rightarrow \frac{\sqrt{13}}{3}-1 \approx 0.20185$, and $S^{trine}_{\max}=\frac{C^{\textrm{trine}}_{\max}}{6} \rightarrow 0.03364 \textrm{ or } 3.36\%$.
\end{theorem}

Thus the quantum probability of anticorrelation can exceed the generalized-noncontextual bound by an amount arbitrarily close to $0.03364$ or about $3.36\%$ for trine spin axes. 
The quantum degree of anti-correlation for this violation is $R_3^Q=S^{trine}_{\max}+\big(1-\frac{\eta}{3}\big)\rightarrow 0.8114$ and the generalized-noncontextual bound is $\big(1-\frac{\eta}{3}\big)\rightarrow\frac{7}{9}\approx 0.7778$.
The proof of Theorem \ref{thm2} follows from Appendix \ref{optimal}.
 
\section{Conclusions}\label{conclude}
A violation of the LSW inequality is interesting primarily because the benchmark for nonclassicality set by
generalized noncontextuality is more stringent than the one set by the traditional notion of KS-noncontextuality. The LSW inequality 
takes into account, for example, the possibility that the measurement apparatus could introduce anticorrelations that have nothing
to do with hidden variable(s) one could associate with the system's preparation. This would allow violation of the KS-noncontextual 
bound of $\frac{2}{3}$ when the measurement is unsharp ($\eta<1$) even though this violation could purely be a result of 
noise coming from elsewhere, such as the measurement apparatus, rather than a consequence of quantum theory. A violation of the LSW inequality, on
the other hand, rules out this possibility and certifies genuine nonclassicality that cannot be attributed to hidden variables
associated with either the preparation or the noise. As argued by Spekkens \cite{genNC}, the appropriate notion of noncontextuality
for unsharp measurements is one that allows outcome-indeterministic response functions.

An interesting open question is whether such a violation is possible in higher dimensional systems and whether the amount of violation could be higher for these than for a qubit.
Whether a state-independent violation of the LSW inequality is possible in higher dimensions also remains an open question.
Our result also hints at the fact that perhaps all contextuality scenarios may be realizable and contextuality demonstrated if we consider the possibilities that general quantum measurements allow.
In particular, scenarios that involve pairwise compatibility between all measurements but no global compatibility may be realizable within quantum theory. Specker's scenario is the simplest such
example we have considered. Indeed, as later shown in Ref. \cite{KHF} after the present work was completed, quantum theory does admit all contextuality scenarios since it allows one to realize any conceivable set of (in)compatibility
relations between a set of observables.

In summary, the joint measurability allowed in a theory restricts the kind of contextuality scenarios that can arise in it. Quantum theory admits Specker's contextuality scenario with unsharp measurements \cite{OS}.
Further, as we have shown, quantum theory allows violation of the LSW inequality in this scenario. Thus quantum theory is contextual even in the simplest contextuality scenario. Whether,
and to what extent, this is the case with more complicated contextuality scenarios realizable, for example, via the construction in Ref. \cite{KHF} remains to be explored.

\emph{Note.---}In Ref. \cite{FyuOh}, which appeared after completion of the present work, the authors deal with the LSW inequality and make some remarks on the results of this paper. We refer the 
interested reader to Ref. \cite{RK} for a discussion of claims in Ref. \cite{FyuOh} compared to the results of this paper. See also Appendix \ref{relaxincomp} 
for a brief remark on the question of triplewise incompatibility.
\section*{Acknowledgments}
R.~K. would like to thank Rob Spekkens for comments on earlier drafts of this work; in particular, for asking whether a state-dependent violation of the LSW inequality was possible, and for 
clarifying the operational meaning of the generalized-noncontextual bound in this inequality. Thanks are also due to Andreas Winter
for asking some tough questions about benchmarks for nonclassicality. We also thank the anonymous referees for their  
comments which made us re-examine and revise some of our results.

\begin{appendix}
\section{Generalized-noncontextual models vs. KS-noncontextual models}\label{modelcompare}
The traditional assumption of KS-noncontextuality entails two things: measurement noncontextuality and outcome-determinism for sharp measurements \cite{genNC}.
Given a set of measurements $\{M_1,\dots,M_N\}$, measurement noncontextuality is the assumption that the response function for each measurement
is insensitive to contexts---jointly measurable subsets---that it may be a part of: $\forall M_i, p(X_i|M_i;\lambda) \in [0,1]$. Here $X_i$ is an 
outcome for measurement $M_i$ and $\lambda$ is the hidden variable associated with the system's preparation. Outcome-determinism
is the further assumption that $\forall M_i, p(X_i|M_i;\lambda) \in \{0,1\}$, i.e., response functions are outcome-deterministic.
A KS-NCHV model is one that makes these two assumptions for sharp (projective) measurements. A KS-inequality is a constraint on 
measurement statistics obtained under these two assumptions. 
A generalized-noncontextual model, on the other hand, derives outcome-determinism for sharp measurements as a consequence of preparation noncontextuality \cite{genNC}.
For unsharp measurements, however, outcome-determinism is not implied by generalized-noncontextuality and one needs to model these 
measurements by outcome-indeterministic response functions \cite{genNC, noODUM}. We refer the reader to Ref. \cite{noODUM} for a detailed critique of 
the assumption of outcome-determinism for unsharp measurements and for arguments on the reasonableness, and generality, of the notion of noncontextuality 
for unsharp measurements that is the basis of our work. Indeed, the qubit effects we need to write the response functions for are of the form: $E^k_{\pm}=\eta \Pi^k_{\pm}+(1-\eta)\frac{I}{2}$.
We will relabel the outcomes according to $\{+1\rightarrow 0, -1\rightarrow 1\}$ so that $X_k \in \{0,1\}$ in what follows.
Liang, Spekkens and Wiseman (LSW) argued \cite{OS} that the response function for these effects in a generalized-noncontextual model
should be $p(X_k|M_k;\lambda)=\eta[X_k(\lambda)]+(1-\eta)\left(\frac{1}{2}[0]+\frac{1}{2}[1]\right)$,
where $p(X)=[x]$ denotes the point distribution given by the Kronecker delta function $\delta_{X,x}$. For $\eta=1$ (sharp measurements) this would be the traditional KS-noncontextual model. When $\eta<1$ (unsharp measurements),
the second ``coin flip'' term in the response function, $(\frac{1}{2}[0]+\frac{1}{2}[1])$, begins to play a role. This term is not conditioned
by $\lambda$, the hidden variable associated with the system's preparation, but is instead the response function for tossing a fair
coin regardless of what measurement is being made. It characterizes the random noise introduced, for example, by the measuring apparatus. 
The important thing to note is that this noise is uncorrelated with the system's hidden variable $\lambda$.

Given these single-measurement response functions, one needs to figure out pairwise response functions for pairwise
joint measurements of the three qubit POVMs. LSW \cite{OS} argued that the pairwise response functions maximizing the average anti-correlation
$R_3$ and consistent with the single-measurement response functions are given by 
\begin{eqnarray}\nonumber
 p(X_i,X_j|M_{ij};\lambda)&=&\eta [X_i(\lambda)][X_j(\lambda)]\\
 &+&(1-\eta)\left(\frac{1}{2}[0][1]+\frac{1}{2}[1][0]\right),
\end{eqnarray}
for all pairs of measurements $(ij)\in \{(12),(13),(23)\}$. This generalized-noncontextual model for these measurements turns out to be 
KS-contextual in the sense that the three pairwise response functions do not admit a joint probability distribution over the three measurement
outcomes, $p(X_1,X_2,X_3|\lambda)$, that is consistent with all three of them. Indeed this LSW-model maximizes the average anticorrelation
possible in Specker's scenario given the single-measurement response functions, thus allowing us to obtain the LSW inequality
$R_3\leq 1-\frac{\eta}{3}$. Let us note the two bounds separately:

\begin{eqnarray}
 R_3 &\leq& R_3^{KS}\equiv\frac{2}{3}\\
 R_3 &\leq& R_3^{LSW}\equiv1-\frac{\eta}{3}
\end{eqnarray}

To be clear, the assumptions leading to the LSW inequality are:
\begin{itemize}
 \item Measurement noncontextuality, 
 \item Outcome-determinism for projective measurements,
 \item No outcome-determinism for nonprojective measurements.
\end{itemize}

On the other hand, the assumptions that lead to the corresponding KS-inequality (upper bound of $\frac{2}{3}$) are:
\begin{itemize}
 \item Measurement noncontextuality,
 \item Outcome-determinism for \emph{all} (projective as well as nonprojective) measurements.
\end{itemize}

The first set of assumptions is clearly weaker than the second set of assumptions and violation of the LSW inequality rules out even this weaker notion of noncontextuality.
Our main result in this paper is that there exists a choice of measurement directions, $\{\hat{n}_1,\hat{n}_2,\hat{n}_3\}$, and
a choice of $\eta$ for some state $\rho$ such that the quantum probability of anticorrelation, $R_3^Q$, beats the generalized-noncontextual
bound $R_3^{LSW}$. This rules out the possibility of being able to generate these correlations by classical means, as in the LSW-model, for at least 
some values of sharpness parameter $\eta$. Of course, if $\eta=0$, then the generalized-noncontextual bound becomes trivial and the
question of violation does not arise---this situation corresponds to the case where for any of the three pairwise joint measurements,
the measuring apparatus outputs one of the two anticorrelated outcomes by flipping a fair coin and there is no functional dependence
of the response function on $\lambda$. In other words, one could generate perfect anti-correlation in a generalized-noncontextual
model if $\eta=0$. However, as long as one is performing a nontrivial measurement (where $\eta>0$) there is a constraint on
the degree of anticorrelation imposed by generalized-noncontextuality. What we establish is that generalized-noncontextuality
cannot account for the degree of anticorrelation observed in quantum theory. Clearly, quantum theory is nonclassical even given a more stringent 
benchmark than the one set by KS-noncontextuality. A violation of the KS-noncontextual bound, $R_3^{KS}$, is possible in a generalized-noncontextual model,
so such a violation is not in itself a signature of nonclassicality. On the other hand, violation of the generalized-noncontextual bound, $R_3^{LSW}$, should 
be considered a signature of genuine nonclassicality in that it isn't attributable either to the system's hidden variable or random noise (from the
measuring apparatus or elsewhere).

\section{Bounds on $\eta$ from joint measurability}\label{bounds}

In Appendix F of \cite{OS}, Theorem 13, the authors obtain necessary and sufficient conditions for joint measurability of noisy spin-1/2 observables. We note, as pointed out by an anonymous referee,
that the claimed necessary condition in the aforementioned theorem is incorrect, while the sufficient condition holds. Here we prove a necessary condition for joint measurability, one we use in the
main text for triplewise joint measurability, by revising the argument for necessity made by Liang et al.:

\begin{theorem}
 Given a set of qubit POVMs, $\{\{E^k_{X_k}: X_k \in \{+1,-1\}\}|k \in \{1\dots N\}\}$, of the form 
\begin{equation}
 E^k_{X_k}=\frac{1}{2}I+\frac{1}{2}\vec{\sigma}.X_k\eta\hat{n}_k,
\end{equation}
and defining $2^N$ 3-vectors
\begin{equation}
 \vec{m}_{X_1\dots X_N}\equiv \sum_{k=1}^{N}X_k\hat{n}_k,
\end{equation}
a necessary condition for all the POVMs to be jointly measurable is that
\begin{equation}
 \eta \leq \frac{1}{N} \max_{X_1\dots X_N} \{|\vec{m}_{X_1\dots X_N|}\},
\end{equation}
and a sufficient condition is that 
\begin{equation}
 \eta \leq \frac{2^N}{\sum_{X_1\dots X_N}|\vec{m}_{X_1\dots X_N}|}.
\end{equation}
\end{theorem}

\textbf{Proof.} We will only prove the necessary condition, which we use in the main text, and refer the reader to Ref.\cite{OS}, Appendix F, for proof of the sufficient condition. Note that 
$\eta=\Tr\left[(\vec{\sigma}.X_k\hat{n}_k)E^k_{X_k}\right]$. Since this holds $\forall X_k,k$, we have
\begin{equation}
 \eta=\frac{1}{2N}\sum_{k=1}^{N}\sum_{X_k}\Tr\left[(\vec{\sigma}.X_k\hat{n}_k)E^k_{X_k}\right]
\end{equation}
If all the POVMs are jointly measurable, then we must necessarily have a joint POVM $\{E_{X_1\dots X_N}\}$ such that 
\begin{equation}
 E^k_{X_k}=\sum_{X_1\dots X_{k-1},X_{k+1}\dots X_N}E_{X_1\dots X_N}.
\end{equation}
Then,
\begin{equation}
 \eta=\frac{1}{2N}\sum_{X_1\dots X_N}\Tr\left[\left(\vec{\sigma}.\sum_{k=1}^{N}X_k\hat{n}_k\right)E_{X_1\dots X_N}\right],
\end{equation}
and using $\hat{m}_{X_1\dots X_N}\equiv\vec{m}_{X_1\dots X_N}/|\vec{m}_{X_1\dots X_N}|$, we have
\begin{equation}
 \eta=\frac{1}{2N}\sum_{X_1\dots X_N}|\vec{m}_{X_1\dots X_N}|\Tr\left[\left(\vec{\sigma}.\hat{m}_{X_1\dots X_N}\right)E_{X_1\dots X_N}\right].
\end{equation}
Further,
\begin{equation}
 \Tr\left[\left(\vec{\sigma}.\hat{m}_{X_1\dots X_N}\right)E_{X_1\dots X_N}\right]\leq \Tr\left[E_{X_1\dots X_N}\right],
\end{equation}
which yields the inequality
\begin{equation}
 \eta \leq \frac{1}{2N}\sum_{X_1\dots X_N}|\vec{m}_{X_1\dots X_N}|\Tr\left[E_{X_1\dots X_N}\right].
\end{equation}
Now, $\sum_{X_1\dots X_N} E_{X_1\dots X_N}=I$, and therefore, 
\begin{equation}
 \sum_{X_1\dots X_N} \frac{1}{2} \Tr\left[E_{X_1\dots X_N}\right]=1.
\end{equation}
Also, $0\leq\frac{1}{2} \Tr\left[E_{X_1\dots X_N}\right]\leq1$, so we have, by convexity of the set $\left\{\frac{1}{2} \Tr\left[E_{X_1\dots X_N}\right]\right\}_{X_1\dots X_N}$,
\begin{equation}
 \eta \leq \frac{1}{N}\max_{X_1\dots X_N}\left\{|\vec{m}_{X_1\dots X_N}|\right\},
\end{equation}
which is a necessary condition for joint measurability.\proofend

For $N=3$ we obtain the necessary condition for triplewise joint measurability which is used in the main text for computing $\eta_l$.
The necessary and sufficient condition for pairwise joint measurability is given by
\begin{equation}
 1+\eta^4(\hat{n}_i.\hat{n}_j)^2-2\eta^2\geq 0, \quad \forall (ij)\in\{(12),(13),(23)\}.
\end{equation}
This is obtained as a special case, for the present problem, of the more general necessary and sufficient condition for joint measurability of unsharp qubit observables obtained in Ref. \cite{YuOh}.
Using $\hat{n}_i.\hat{n}_j=\cos \theta_{ij}$, this inequality becomes
\begin{equation}
 \left(\eta^2-\frac{1}{1-|\sin \theta_{ij}|}\right)\left(\eta^2-\frac{1}{1+|\sin \theta_{ij}|}\right)\geq 0.
\end{equation}
Since $0\leq \eta \leq1$, the necessary and sufficient condition for pairwise joint measurability becomes
\begin{equation}
 \eta \leq \min_{(ij)\in\{(12),(23),(13)\}} \left\{ \frac{1}{\sqrt{1+|\sin \theta_{ij}|}} \right\},
\end{equation}
which is used to compute $\eta_u$ in the main text.

{\it Orthogonal spin axes:} $\hat{n}_i.\hat{n}_j=0 \quad \forall (ij) \in \{(12),(13),(23)\}$. The necessary and sufficient joint measurability condition is 
\begin{equation}
 \frac{1}{\sqrt{3}}<\eta\leq\frac{1}{\sqrt{2}}
\end{equation}
\\
{\it Trine spin axes:} $\hat{n}_i.\hat{n}_j=-1/2 \quad \forall (ij) \in \{(12),(13),(23)\}$. The necessary and sufficient joint measurability condition is
\begin{equation}
 \frac{2}{3}<\eta\leq\sqrt{3}-1
\end{equation}

\section{Constructing the joint measurement POVM}\label{jointconstruct}
The joint measurement POVM $G_{ij}$ for $\{M_i,M_j\}$ should satisfy the marginal condition:
\begin{eqnarray}\label{marg1}
G^{ij}_{++}+G^{ij}_{+-}=E^i_+,\quad
G^{ij}_{-+}+G^{ij}_{--}=E^i_-,\\
G^{ij}_{++}+G^{ij}_{-+}=E^j_+,\quad
G^{ij}_{+-}+G^{ij}_{--}=E^j_-.\label{marg2}
\end{eqnarray}
Also, the joint measurement should consist of valid effects:
\begin{equation}\label{valid}
0\leq G^{ij}_{++},G^{ij}_{+-},G^{ij}_{-+},G^{ij}_{--} \leq I,
\end{equation}
where $I$ is the $2\times2$ identity matrix. The general form of the joint measurement effects is:
\begin{eqnarray}
G^{ij}_{++}&=&\frac{1}{2}[\frac{\alpha_{ij}}{2}I+\vec{\sigma}.\vec{a}^{ij}_{++}],\\
G^{ij}_{+-}&=&\frac{1}{2}[(1-\frac{\alpha_{ij}}{2})I+\vec{\sigma}.\vec{a}^{ij}_{+-}],\\
G^{ij}_{-+}&=&\frac{1}{2}[(1-\frac{\alpha_{ij}}{2})I+\vec{\sigma}.\vec{a}^{ij}_{-+}],\\
G^{ij}_{--}&=&\frac{1}{2}[\frac{\alpha_{ij}}{2}I+\vec{\sigma}.\vec{a}^{ij}_{--}],
\end{eqnarray}
where each effect is parameterized by four real numbers. From the marginal condition, eqs. (\ref{marg1}-\ref{marg2}), it follows that:
\begin{eqnarray}\label{bega}
\vec{a}^{ij}_{++}+\vec{a}^{ij}_{+-}=\eta \hat{n}_i,\quad
\vec{a}^{ij}_{-+}+\vec{a}^{ij}_{--}=-\eta \hat{n}_i,\\
\vec{a}^{ij}_{-+}+\vec{a}^{ij}_{++}=\eta \hat{n}_j,\quad
\vec{a}^{ij}_{--}+\vec{a}^{ij}_{+-}=-\eta \hat{n}_j.\label{begb}
\end{eqnarray}
These can be rewritten as:
\begin{eqnarray}\label{beg1}
2\vec{a}^{ij}_{++}+\vec{a}^{ij}_{+-}+\vec{a}^{ij}_{-+}&=&\eta (\hat{n}_i+\hat{n}_j),\\
2\vec{a}^{ij}_{+-}+\vec{a}^{ij}_{++}+\vec{a}^{ij}_{--}&=&\eta (\hat{n}_i-\hat{n}_j),\\
2\vec{a}^{ij}_{-+}+\vec{a}^{ij}_{++}+\vec{a}^{ij}_{--}&=&\eta (-\hat{n}_i+\hat{n}_j),\\
2\vec{a}^{ij}_{--}+\vec{a}^{ij}_{+-}+\vec{a}^{ij}_{-+}&=&\eta (-\hat{n}_i-\hat{n}_j).\label{end1}
\end{eqnarray}
From eqs. (\ref{bega}-\ref{begb}) it follows that:
$$(\vec{a}^{ij}_{++}+\vec{a}^{ij}_{--})+(\vec{a}^{ij}_{-+}+\vec{a}^{ij}_{+-})=0.$$
So one can define:
$$\vec{a}_{ij}\equiv\vec{a}^{ij}_{+-}+\vec{a}^{ij}_{-+} \Rightarrow \vec{a}^{ij}_{++}+\vec{a}^{ij}_{--}=-\vec{a}_{ij}.$$
Now, from eqs. (\ref{beg1})-(\ref{end1}) the following are obvious:
\begin{eqnarray}
\vec{a}^{ij}_{++}&=&\frac{1}{2}[\eta(\hat{n}_i+\hat{n}_j)-\vec{a}_{ij}],\\
\vec{a}^{ij}_{+-}&=&\frac{1}{2}[\eta(\hat{n}_i-\hat{n}_j)+\vec{a}_{ij}],\\
\vec{a}^{ij}_{-+}&=&\frac{1}{2}[\eta(-\hat{n}_i+\hat{n}_j)+\vec{a}_{ij}],\\
\vec{a}^{ij}_{--}&=&\frac{1}{2}[\eta(-\hat{n}_i-\hat{n}_j)-\vec{a}_{ij}].
\end{eqnarray}
This gives the general form of the joint measurement POVMs in the main text. For qubit effects, $G^{ij}_{X_iX_j}$, where $X_i, X_j \in \{+1,-1\}$, the valid effect condition (\ref{valid}) is equivalent to the following \cite{heinosaari}:
\begin{eqnarray}
 |\vec{a}^{ij}_{++}| &\leq& \frac{\alpha_{ij}}{2} \leq 2-|\vec{a}^{ij}_{++}|,\\
 |\vec{a}^{ij}_{+-}| &\leq& 1-\frac{\alpha_{ij}}{2} \leq 2-|\vec{a}^{ij}_{+-}|,\\
 |\vec{a}^{ij}_{-+}| &\leq& 1-\frac{\alpha_{ij}}{2} \leq 2-|\vec{a}^{ij}_{-+}|,\\
 |\vec{a}^{ij}_{--}| &\leq& \frac{\alpha_{ij}}{2} \leq 2-|\vec{a}^{ij}_{--}|.
\end{eqnarray}
These inequalities can be combined and rewritten as:
\begin{equation}
 2 \max \{|\vec{a}^{ij}_{++}|,|\vec{a}^{ij}_{--}|\}\leq\alpha_{ij}\leq 2-2 \max \{|\vec{a}^{ij}_{+-}|,|\vec{a}^{ij}_{-+}|\}
\end{equation}
where 
\begin{eqnarray}\nonumber
\max\{|\vec{a}^{ij}_{++}|,|\vec{a}^{ij}_{--}|\} \\ =\sqrt{\frac{\eta^2}{2}(1+\hat{n}_i.\hat{n}_j)+\frac{|\vec{a}_{ij}|^2}{4}+\frac{\eta}{2}|(\hat{n}_i+\hat{n}_j).\vec{a}_{ij}|}\nonumber
\end{eqnarray}
and
\begin{eqnarray}\nonumber
\max \{|\vec{a}^{ij}_{+-}|,|\vec{a}^{ij}_{-+}|\} \\ =\sqrt{\frac{\eta^2}{2}(1-\hat{n}_i.\hat{n}_j)+\frac{|\vec{a}_{ij}|^2}{4}+\frac{\eta}{2}|(\hat{n}_i-\hat{n}_j).\vec{a}_{ij}|}.\nonumber
\end{eqnarray}
This is the condition for a valid joint measurement used in inequalities (\ref{valid1}-\ref{valid2}) in the main text.
\section{Optimal state-dependent violation for measurements in a plane}\label{optimal}
We need to maximize $C\equiv 2\eta-(\sum_{(ij)}\alpha_{ij}-|\vec{a}|)$ to obtain the optimal violation of the LSW inequality. Subject to satisfaction of the joint measurability constraints (13-14) in the main text,
we have
\begin{eqnarray}\nonumber
C_{max}&=&\max_{\{\hat{n}_1, \hat{n}_2, \hat{n}_3\},\{\vec{a}_{ij}\}, \eta} \{2\eta+|\vec{a}|-\sum_{(ij)}\alpha_{ij}\}\\\nonumber
&\leq& \max_{\{\hat{n}_1, \hat{n}_2, \hat{n}_3\},\{\vec{a}_{ij}\},\eta} \big\{2\eta+\sum_{(ij)}|\vec{a}_{ij}|\\&-&\sum_{(ij)}\sqrt{2\eta^2(1+\hat{n}_i.\hat{n}_j)+|\vec{a}_{ij}|^2}\big\}\\\nonumber
\end{eqnarray}
The inequality above follows from the fact that
\begin{equation}
 |\vec{a}|=\sqrt{\sum_{(ij)}|\vec{a}_{ij}|^2+2(\vec{a}_{12}.\vec{a}_{13}+\vec{a}_{12}.\vec{a}_{23}+\vec{a}_{13}.\vec{a}_{23})},
\end{equation}
so that $|\vec{a}|\leq \sum_{(ij)}|\vec{a}_{ij}|$, and 
\begin{eqnarray}\nonumber
 \sum_{(ij)}\sqrt{2\eta^2(1+\hat{n}_i.\hat{n}_j)+|\vec{a}_{ij}|^2}\\\nonumber \leq \sum_{(ij)}\sqrt{2\eta^2(1+\hat{n}_i.\hat{n}_j)+|\vec{a}_{ij}|^2+2\eta|(\hat{n}_i+\hat{n}_j).\vec{a}_{ij}|}\\\nonumber
 \leq \sum_{(ij)}\alpha_{ij}.\nonumber
\end{eqnarray}
Also, we have
\begin{eqnarray}\nonumber
 \sum_{(ij)}\sqrt{2\eta^2(1+\hat{n}_i.\hat{n}_j)+|\vec{a}_{ij}|^2}\\\nonumber\geq \sum_{(ij)}\sqrt{2\eta^2(1+\hat{n}_i.\hat{n}_j)+|\vec{a}_{ij}|^2}\big|_{\text{coplanar}, \phi_3=\pi} 
\end{eqnarray}
That is, for a fixed $|\vec{a}_{ij}|$, $\sum_{(ij)}\sqrt{2\eta^2(1+\hat{n}_i.\hat{n}_j)+|\vec{a}_{ij}|^2}$ is smallest when the measurement directions
$\{\hat{n}_1, \hat{n}_2, \hat{n}_3\}$ are coplanar and $\phi_3=\pi$. From eqs. (\ref{mmts1}-\ref{mmts2}) in the main text, $\hat{n}_2.\hat{n}_3=\cos \theta_{23}=\sin \theta_{12}\sin \theta_{13} \cos \phi_3+\cos \theta_{12}\cos \theta_{13}$.
When $\phi_3=0 \textrm{ or }\pi$, the three measurements are coplanar and there are only two free angles, $\hat{n}_1.\hat{n}_2=\cos \theta_{12}$
and $\hat{n}_1.\hat{n}_3=\cos \theta_{13}$, while the third angle is fixed by these two: $\hat{n}_2.\hat{n}_3=\cos \theta_{23}=\cos (\theta_{12}-\theta_{13}) \textrm{ or }\cos (\theta_{12}+\theta_{13})$. Since 
$\cos(\theta_{12}+\theta_{13})\leq \cos(\theta_{23}) \leq \cos(\theta_{12}-\theta_{13})$, for any given $\theta_{12}$ and $\theta_{13} \in (0,\pi)$, $\cos \theta_{23}$ is smallest when
$\phi_3=\pi$. Hence, we choose the three measurements to be coplanar such that $\phi_3=\pi$ and $\cos \theta_{23}=\cos (\theta_{12}+\theta_{13})$. Any other choice of $\{\hat{n}_1, \hat{n}_2, \hat{n}_3\}$
will give a larger value of $\cos \theta_{23}$, hence also $\sum_{(ij)}\sqrt{2\eta^2(1+\hat{n}_i.\hat{n}_j)+|\vec{a}_{ij}|^2}$. So,
\begin{eqnarray}\nonumber
C_{max}&\leq& \max_{\hat{n}_1.\hat{n}_2, \hat{n}_1.\hat{n}_3, \{|\vec{a}_{ij}|\}, \eta} \big\{2\eta+\sum_{(ij)}|\vec{a}_{ij}|\\\nonumber&-&\sum_{(ij)}\sqrt{2\eta^2(1+\hat{n}_i.\hat{n}_j)+|\vec{a}_{ij}|^2}\big|_{\text{coplanar}, \phi_3=\pi}\big\}\\\nonumber
\end{eqnarray}
We will now argue that this inequality for $C_{max}$ can be replaced by an equality. Let us take coplanar measurement directions $\{\hat{n}_1, \hat{n}_2, \hat{n}_3\}$ such that $\phi_3=\pi$.
We also take all the $\vec{a}_{ij}$ parallel to each other, i.e., $\vec{a}_{12}.\vec{a}_{13}=|\vec{a}_{12}||\vec{a}_{13}|$, $\vec{a}_{12}.\vec{a}_{23}=|\vec{a}_{12}||\vec{a}_{23}|$,
and $\vec{a}_{13}.\vec{a}_{23}=|\vec{a}_{13}||\vec{a}_{23}|$, so that $|\vec{a}|=|\vec{a}_{12}|+|\vec{a}_{13}|+|\vec{a}_{23}|$. Besides, $|(\hat{n}_i+\hat{n}_j).\vec{a}_{ij}|=0$ $\forall (ij)\in \{(12),(13),(23)\}$.
From these conditions it follows that each $\vec{a}_{ij}$ is perpendicular to the plane and 
$\forall (ij) \in \{(12),(13),(23)\}$, $\vec{a}_{ij}.\hat{n}_i=\vec{a}_{ij}.\hat{n}_j=0$. This allows us to choose $\alpha_{ij}=\sqrt{2\eta^2(1+\hat{n}_i.\hat{n}_j)+|\vec{a}_{ij}|^2}$.
So, in our optimal configuration, the measurement directions are coplanar while the $\vec{a}_{ij}$'s are parallel to each other and perpendicular to the plane of measurements. Note that this also means $\vec{a}$ will be parallel to $\vec{a}_{ij}$ and therefore perpendicular to
the plane of measurements, and so will be the optimal state (which is parallel to $\vec{a}$). With these optimality conditions satisfied, the optimal violation can now be written as
\begin{eqnarray}
 C_{max}&=&\max_{\hat{n}_1.\hat{n}_2, \hat{n}_1.\hat{n}_3, \{|\vec{a}_{ij}|\},\eta}\big\{2\eta+\sum_{(ij)}\big(|\vec{a}_{ij}|\\&-&\sqrt{2\eta^2(1+\hat{n}_i.\hat{n}_j)+|\vec{a}_{ij}|^2}\big)\big\}.
\end{eqnarray}
The constraints from joint measurability (13-14) become
\begin{equation}
|\vec{a}_{ij}|\leq \sqrt{1+\eta^4(\hat{n}_i.\hat{n}_j)^2-2\eta^2}.
\end{equation}
Now,
\begin{eqnarray}\nonumber
C_{max}&\leq&\max_{\hat{n}_1.\hat{n}_2, \hat{n}_1.\hat{n}_3, \{|\vec{a}_{ij}|\},\eta}\big\{2\eta+\\\nonumber&&\sum_{(ij)}\big(\sqrt{1+\eta^4(\hat{n}_i.\hat{n}_j)^2-2\eta^2}-(1+\eta^2 \hat{n}_i.\hat{n}_j)\big)\big\}.
\end{eqnarray}
The upper bound follows from the fact that $f(x,y)=x-\sqrt{x^2+2\eta^2(1+y)}$, where $0\leq x \leq \sqrt{1+\eta^4y^2-2\eta^2}$ and $-1<y<1$, is an increasing function of 
$x$ for a fixed $y$, i.e., $(\frac{\partial{f}}{\partial{x}})_y>0$. Here $x\equiv|\vec{a}_{ij}|$ and $y\equiv \hat{n}_i.\hat{n}_j$. So, taking $|\vec{a}_{ij}|=\sqrt{1+\eta^4(\hat{n}_i.\hat{n}_j)^2-2\eta^2}$,
we have
\begin{eqnarray}\nonumber
 &&C_{max}^{\{\hat{n}_i\},\eta}\\\nonumber&\equiv& 2\eta+\sum_{(ij)}\big(\sqrt{1+\eta^4(\hat{n}_i.\hat{n}_j)^2-2\eta^2}-(1+\eta^2 \hat{n}_i.\hat{n}_j)\big)\\
\end{eqnarray}
Note that $\alpha_{ij}=1+\eta^2 \hat{n}_i.\hat{n}_j$ for $|\vec{a}_{ij}|=\sqrt{1+\eta^4(\hat{n}_i.\hat{n}_j)^2-2\eta^2}$. $C_{max}^{\{\hat{n}_i\},\eta}$ 
is the maximum value of $C$ for a given coplanar choice of measurement directions $\{\hat{n}_1,\hat{n}_2,\hat{n}_3\}$ and sharpness parameter $\eta$. 
\section{Note on triplewise incompatibility}\label{relaxincomp}
We have worked with the constraint of triplewise incompatibility originally employed by LSW \cite{OS}. The intuition behind this constraint was to ensure that the binary qubit POVMs we consider
do not trivially admit a joint distribution, in which case there should be no contextuality even with respect to the KS-inequality bound of $2/3$. However, the situation in respect of POVMs 
turns out to be richer than expected: it is possible to construct pairwise joint measurements of these qubit POVMs such that they violate the LSW inequality, and it is also possible for these qubit POVMs
to be triplewise jointly measurable. It's just that any triplewise joint measurement that one may construct for these qubit POVMs will not marginalize to pairwise joint measurements capable of
violating the LSW inequality (or even the KS-inequality for this scenario). However, it may still be possible to construct pairwise joint measurements for the qubit POVMs that do not arise as marginals
of any triplewise joint measurement and can therefore violate the LSW inequality.

On relaxing the requirement of triplewise incompatibility in our optimal (trine axes) scenario, $\eta>\eta_l=\frac{2}{3}$, we find that the maximum violation of the LSW inequality occurs at 
$\eta=0.4566$, with $R_3^Q=0.9374$ and $R_3^{LSW}=0.8478$, the violation being $0.0896$ or about $8.96\%$. This is a straightforward consequence of Eq. (\ref{cmax}) in the main text. This result has 
been quoted in Theorem 3 of Ref. \cite{FyuOh}, where the peculiar behaviour of POVMs with respect to joint measurability has also been discussed. We note, however, that the claim in Ref. \cite{FyuOh} that \emph{``LSW's inequality can be regarded as a genuine KS inequality''}
is fallacious, which should be clear from our discussion of the LSW inequality and its comparison with the corresponding KS-inequality in Appendix \ref{modelcompare}.
For further related discussion, see Ref. \cite{RK}.
\end{appendix}
\end{document}